\documentclass[fleqn, a4paper, oneside]{imsart}

\usepackage{amsthm,amsmath,amssymb}
\usepackage{graphicx}
\usepackage{makeidx}
\usepackage[authoryear]{natbib}

\bibpunct{(}{)}{;}{a}{,}{,}
\RequirePackage[OT1]{fontenc}

\startlocaldefs

\newtheorem{Lemma}{Lemma}
\newenvironment{Proof}{\noindent{\bf Proof:}}{\hfill\rule{2mm}{2mm}\vspace{0.3cm}}

\theoremstyle{remark}

\theoremstyle{definition}

\def\P{\mathbb{P}}
\def\m{\mathrm{m}}

\def\GG{\mathcal{G}}

\endlocaldefs

\begin{document}

\begin{frontmatter}


\title{Computationally efficient algorithms for statistical image processing. Implementation in R}
\runtitle{Image processing in R.}


\begin{aug}
\author{\snm{Mikhail} \fnms{Langovoy}
\ead[label=e1]{langovoy@eurandom.tue.nl}}

\affiliation{
        EURANDOM,\\
         The Netherlands.}

\address{Mikhail Langovoy, Technische Universiteit Eindhoven, \\
EURANDOM, P.O. Box 513,
\\
5600 MB, Eindhoven, The Netherlands\\
\printead{e1}\\
Phone: (+31) (40) 247 - 8113\\
Fax: (+31) (40) 247 - 8190\\}

\and

\author{\snm{Olaf} \fnms{Wittich} \ead[label=e2]{o.wittich@tue.nl}}\corref{}\thanksref{t2}

\affiliation{
        Technische Universiteit Eindhoven and EURANDOM,\\
         The Netherlands.}

\address{Olaf Wittich, Technische Universiteit Eindhoven and \\
EURANDOM, P.O. Box 513,
\\
5600 MB, Eindhoven, The Netherlands\\
\printead{e2}\\
Phone: (+31) (40) 247 - 2499}

\thankstext{t2}{Corresponding author.}

\runauthor{M. Langovoy and O. Wittich}
\end{aug}

\begin{abstract}
In the series of our earlier papers on the subject, we proposed a novel statistical hypothesis testing method for detection of objects in noisy images. The method uses results from percolation theory and random graph theory. We developed algorithms that allowed to detect objects of unknown shapes in the presence of nonparametric noise of unknown level and of unknown distribution. No boundary shape constraints were imposed on the objects, only a weak bulk condition for the object's interior was required. Our algorithms have linear complexity and exponential accuracy. In the present paper, we describe an implementation of our nonparametric hypothesis testing method. We provide a program that can be used for statistical experiments in image processing. This program is written in the statistical programming language R.\\
\end{abstract}


\begin{keyword}
\kwd{Image analysis} \kwd{signal detection} \kwd{image reconstruction} \kwd{percolation} \kwd{noisy image} \kwd{unsupervised machine learning} \kwd{spatial statistics}
\end{keyword}

\end{frontmatter}

\section{Introduction}\label{Section1}

Assume we observe a noisy digital image on a screen of $N \times N$ pixels. Object detection and image reconstruction for noisy images are two of the cornerstone problems in image analysis. In the series of papers \cite{langovoy_report_2009-035}, \cite{langovoy_davies_wittich}, \cite{Langovoy_Wittich_Square} and \cite{langovoy_report_Robust_Detection}, we proposed a new efficient technique for quick detection of objects in noisy images. Our approach uses mathematical percolation theory.

Detection of objects in noisy images is the most basic problem of image analysis. Indeed, when one looks at a noisy image, the first question to ask is whether there is any object at all. This is also a primary question of interest in such diverse fields as, for example, cancer detection (\cite{Cancer_Detection_1}), automated urban analysis (\cite{Road_Detection_IEEE}), detection of cracks in buried pipes (\cite{Sinha200658}), and other possible applications in astronomy, electron microscopy and neurology. Moreover, if there is just a random noise in the picture, it doesn't make sense to run computationally intensive procedures for image reconstruction for this particular picture. Surprisingly, the vast majority of image analysis methods, both in statistics and in engineering, skip this stage and start immediately with image reconstruction.

The crucial difference of our method is that we do not impose any shape or smoothness assumptions on the \emph{boundary} of the object. This permits the detection of nonsmooth, irregular or disconnected objects in noisy images, under very mild assumptions on the object's interior. This is especially suitable, for example, if one has to detect a highly irregular non-convex object in a noisy image. This is usually the case, for example, in the aforementioned fields of automated urban analysis, cancer detection and detection of cracks in materials. Although our detection procedure works for regular images as well, it is precisely the class of irregular images with unknown shape where our method can be very advantageous.


We approached the object detection problem as a hypothesis testing problem within the class of statistical inverse problems in spatial statistics. We were able to extend our approach to the nonparametric case of unknown noise density, and this density was not assumed smooth or even continuous. It is even possible that the noise distribution doesn't have a density or that the noise distribution is heavy-tailed.

We have shown that there is a deep connection between the spatial structure chosen for the discretisation of the image, the type of the noise distribution on the image  and statistical properties of object detection. These results seem to be of independent interest for the field of spatial statistics.

In our previous papers, we considered the case of square lattices in \cite{langovoy_report_2009-035} and \cite{Langovoy_Wittich_Square}, triangular lattices in \cite{langovoy_davies_wittich} and \cite{langovoy_report_Robust_Detection} and even the case of general periodic lattices in \cite{langovoy_report_Robust_Detection}. In all those cases, we proved that our detection algorithms have linear complexity in terms of the number of pixels on the screen. These procedures are not only asymptotically consistent, but on top of that they have accuracy that grows exponentially with the "number of pixels" in the object of detection. All of the detection algorithms have a built-in data-driven stopping rule, so there is no need in human assistance to stop an algorithm at an appropriate step.

In view of the above, our method can be viewed as an unsupervised learning method, in the language of machine learning. This makes our results valuable for the field of  machine learning as well. Indeed, we do not only propose an unsupervised method, but also prove the method's consistency and even go as far as to prove the rates of convergence.


In this paper, we describe an algorithmic implementation of our nonparametric hypothesis testing / image processing method. We also provide an actual program that was used for our research at the initial stages. This program is written in the statistical programming language R. It is known that, as a programming language, R is rather slow. At present, we have implemented our algorithms in C++, and they run substantially faster. However, R is a free and convenient language for statistical programming, since R has many advanced statistical procedures already built-in. This language also has convenient tools for generating random variables of various types, therefore R is suitable for statistical experiments with our method.

The paper is organized as follows. Section 2.1 describes some basic methods for converting, storing and processing images in R. Our implementation of lattice  neighborhood structures is given in Section 2.2. The Depth First Search algorithm and its R-implementation for image analysis are described in Section 3. In Section 4 we explain how the fast Newman-Ziff algorithm from discrete algorithmic probability can be effectively used in our image analysis method. Subsection 4 describes the theory behind the Newman-Ziff approach, while Subsection 4.1 introduces our modification of this algorithm, suitable for percolation probabilities that are inhomogeneous over the lattice. This modification allows for finer simulation studies for any given finite screen size, and this is particularly useful for small images. Subsections 4.2 and 4.3 are devoted to R-implementations of the Newman-Ziff algorithm and the modified Newman-Ziff algorithms respectively.



\section[Image processing in R]{General remarks about image processing in R}

\subsection{Converting, storing, and processing images}

Image processing in R uses the package {\tt rimage}\index{rimage} which allows to convert an actual picture (in {\tt .jpg}-format) to an R-object called {\tt imagematrix}. In the sequel, we store images as {\em vectors} of grey-values where the entries of the vector are intensities of the corresponding pixels going through the image row by row. The vectors are manipulated and converted to imagematrices which can finally be displayed.


\begin{verbatim}
## SOME CLUSTER MACROS
#
## use only with "library(rimage)"

## Some functions converting data structures
#
#
# the functions convert vectors in matrices and imagematrices;
# they work properly only if the components are either 0 or 1
#
## Function: Image to Matrix (img2mat)

img2mat <- function(img)
{
mat <- matrix(img,nrow(img),ncol(img))
return(mat)
}

## Function: Matrix to Image (mat2img)

mat2img <- function(mat)
{
img <- imagematrix(mat)
return(img)
}

## Function: Matrix to Vector (mat2vec)

mat2vec <- function(mat)
{
vec <- as.vector(mat)
return(vec)
}

## Function: Vector to Matrix (vec2mat)

vec2mat <- function(vec,n,m)
{
mat <- matrix(vec,n,m)
return(mat)
}
\end{verbatim}
The following code just provides you with an example of generating a graphical representation of site percolation on a lattice:
\begin{verbatim}
## EXAMPLE: Percolation
# simulates site? percolation on an n x n lattice
# with parameter 0 <= p <= 1

perco <- function(p,n)
{
v <- rbinom(n*n,1,p)
ma <- vec2mat(v,n)
im <- mat2img(ma)
return(im)
}
\end{verbatim}

\subsection{Neighborhood structures}

For the percolation cluster analysis, it is important to implement the topology of the lattice, i.e. the neighborhood structure. Since the image - i.e. the lattice site intensities - are stored as a vector, the neighborhood structures are stored as {\tt list}-objects. Assigned to every position in the vector (i.e. every {\em site}, the {\tt list}-index) there is a vector of other vector positions, i.e. the indices of the neighboring sites. These lists depend on the size of the rectangular grid and have to be computed only once. Below we give an implementation for the standard cases of 4-, 6-, and 8-neighborhoods on a rectangular grid.


\begin{verbatim}
###### Function: Adjacency Matrix (adj_mat)
#
#
# the function generates a list of all neighbors
# of points in a flat rectangular grid consisting of
# n:columns, m :rows; it works properly only for
# for n,m > 2 !

## 4-neighborhood

adj4mat <- function(n,m)
{

neighbors <- list()

## corners

neighbors[[1]]           <- c(2,n+1)
neighbors[[n]]           <- c(n-1,2*n)
neighbors[[n*m]]         <- c(n*(m-1),n*m -1)
neighbors[[(m-1)*n + 1]] <- c((m-1)*n + 2,(m-2)*n+1)

## boundary

# (a) above

for(i in 2:(n-1)){neighbors[[i]]<- c(i-1,i+1,i+n)}

# (b) below

for(i in ((m-1)*n+2):(n*m-1)){neighbors[[i]]<- c(i-1,i+1,i-n)}

### (c) left

for(i in 1:(m-2)){neighbors[[n*i+1]]<- c(n*(i-1)+1,n*(i+1)+1,
                                                    n*i+2)}

### (d) right

for(i in 2:(m-1)){neighbors[[n*i]]<- c(n*(i-1),n*(i+1),n*i-1)}

## interior

for(i in 1:(m-2))
{
for(j in 1:(n-2))
{
neighbors[[i*n + 1 + j]]<- c(i*n + j,i*n + 2 + j,(i-1)*n+1+j,
                                (i+1)*n+1+j)
}
}

return(neighbors)

}

## 8-neighborhood

adj8mat <- function(n,m)
{

neighbors <- list()

## corners

neighbors[[1]]           <- c(2,n+1,n+2)
neighbors[[n]]           <- c(n-1,2*n,2*n-1)
neighbors[[n*m]]         <- c(n*(m-1),n*m -1,(m-1)*n-1)
neighbors[[(m-1)*n + 1]] <- c((m-1)*n + 2,(m-2)*n+1,(m-2)*n+2)

## boundary

# (a) above

for(i in 2:(n-1)){neighbors[[i]]<- c(i-1,i+1,i+n,i+n-1,i+n+1)}

# (b) below

for(i in ((m-1)*n+2):(n*m-1)){neighbors[[i]]<- c(i-1,i+1,i-n,
                                                    i-n-1,i-n+1)}

### (c) left

for(i in 1:(m-2)){neighbors[[n*i+1]] <- c(n*(i-1)+1,n*(i-1)+2,
                                      n*(i+1)+1,n*(i+1)+2,n*i+2)}

### (d) right

for(i in 2:(m-1)){neighbors[[n*i]]<- c(n*(i-1), n*(i-1)-1,n*(i+1),
                                                n*(i+1)-1,n*i-1)}

## interior

for(i in 1:(m-2))
{
for(j in 1:(n-2))
{
neighbors[[i*n + 1 + j]]<- c(i*n + j,i*n + 2 + j,(i-1)*n+1+j,
        (i-1)*n+j,(i-1)*n+j+2, (i+1)*n+1+j, (i+1)*n+2+j, (i+1)*n+j)
}
}

return(neighbors)

}


## 6-neighborhood (triangular lattice)

adj6mat <- function(n,m)
{

neighbors <- list()

## corners

neighbors[[1]]           <- c(2,n+1,n+2)
neighbors[[n]]           <- c(n-1,2*n,2*n-1)
neighbors[[n*m]]         <- c(n*(m-1),n*m -1)
neighbors[[(m-1)*n + 1]] <- c((m-1)*n + 2,(m-2)*n+1,(m-2)*n+2)

## boundary

# (a) above

for(i in 2:(n-1)){neighbors[[i]]<- c(i-1,i+1,i+n,i+n-1)}

# (b) below

for(i in ((m-1)*n+2):(n*m-1)){neighbors[[i]]<- c(i-1,i+1,i-n,
                                                        i-n+1)}

### (c) left

for(i in 1:(m-2)){neighbors[[n*i+1]] <- c(n*(i-1)+1,n*(i-1)+2,
                                                n*(i+1)+1,n*i+2)}

### (d) right

for(i in 2:(m-1)){neighbors[[n*i]]<- c(n*(i-1),n*(i+1),n*(i+1)-1,
                                            n*i-1)}

## interior

for(i in 1:(m-2))
{
for(j in 1:(n-2))
{
neighbors[[i*n + 1 + j]]<- c(i*n + j,i*n + 2 + j,(i-1)*n+1+j,
                            (i-1)*n+j+2, (i+1)*n+1+j,  (i+1)*n+j)
}
}

return(neighbors)

}
\end{verbatim}


\section[Tarjan's Depth First Search]{Tarjan's Depth First Search algorithm}\label{DFS}

In order to be able to perform statistical analysis of percolation clusters, we need an algorithm that finds those clusters in the thresholded binary image. For this purpose, we used the {\em Depth First Search algorithm} (\cite{Tar:72}). It is remarkable that this algorithm has linear complexity, in terms of the size of the input data. In our particular problem, the algorithm finds the clusters in a number of steps that is linear with respect to the number of pixels on the screen.

The key point of the Depth First Search algorithm is the construction of a so called {\em spanning tree} of a given cluster. Starting from a {\em seed point} with label 1 that belongs to the cluster, we begin to explore all neighbors in an order determined by the ordering of the corresponding vector given in the list that encodes the neighborhood structure. The first neighboring site which belongs to the cluster gets label 2 and we repeat the procedure with this site. The labels are stored. Given we have arrived after some steps at a site with label $n$, the procedure is continued as follows
\begin{enumerate}
\item Explore the neighbors of the current site:
\begin{itemize}
\item The first neighbor that is so far unlabeled and belongs to the cluster gets label $n+1$, we store the current site (with label $n$) as {\em mother}-site of the new site and we proceed with the new site as current site.
\item If there is no neighbor which is unlabeled so far and belongs to the cluster, we return to the {\em mother}-site ({\em backtracking}) as current site.
\end{itemize}
\item The algorithm terminates, if all neighbors of the {\em seed point} are explored.
\end{enumerate}
The cluster is now given by all labeled sites which form a tree (i.e. the {\em spanning tree}) whose root is the {\em seed point} and the different labels display the distance of a given site to the root.\\

\noindent{\bf Remark on running time.} Please note that the subsequent R-routines are extremely slow. A corresponding C-program called from R yields much better results.


\subsection{R-Code: Depth First Search}\label{DFSSIM}

The function that generates a spanning tree for the cluster with seed point $r$ in a given configuration $x\in \lbrace 0,1\rbrace^{a\times b}$ is given by
\begin{verbatim}
## Function: Spanning Tree (spanning_tree)

spanning_tree <- function(x,a,b,r)
{

###### DEPTH FIRST SEARCH

n <- a
m <- b
mother <- c()
flg <- 0

rank <- x
ngh <- adj6mat(n,m)

## starting point

start <- r

current_rnk <- 1
current_pos <- start

rank[start] <- current_rnk

############

while(current_rnk > 0)
{


L <- length(ngh[[current_pos]])
i <- 1

#### check neighbors

while(i <= L)
{
    if(rank[ngh[[current_pos]][i]] != 0){i <- i+1}
    else{## mother (new vertex found)

          mother[ngh[[current_pos]][i]] <- current_pos
          current_pos <- ngh[[current_pos]][i]
          current_rnk <- current_rnk + 1
          rank[current_pos] <- current_rnk
          i <- L + 1
          flg <- 1 ## found something new
        }
}

### nothing found ??

if(flg == 0)
{
### backtracking

current_pos <- mother[current_pos]
current_rnk <- current_rnk - 1
}
else
{flg <- 0}

}

return(rank)
}
\end{verbatim}
The function returns a vector of zeros for points that do not belong to the cluster and numbers $1,2,...$ of labels attached to the cluster points by the Depth First Search algorithm.


\section{The Newman - Ziff algorithm}\label{NZ}

The Newman - Ziff algorithm (\cite{NewZif:00}) is an algorithm to simulate the distribution of (functions of) the size of the largest cluster in percolation theory. In our image analysis method, we use this algorithm to study the null hypothesis, to find critical cluster sizes for object detection and to estimate the power of our test. In addition, this algorithm is extremely useful for controlling the false detection rate and for studying the finite sample performance of our method.

We will explain the Newman - Ziff approach for {\em site percolation} since only minor changes have to be made to cover {\em bond percolation} as well. Let now $\GG$ be a finite graph with $S = \vert\GG\vert$ sites and $0\leq p\leq 1$ the probability that an individual site is {\em active}. Two sites $s,s'\in \GG$ are {\em neighbors} if the graph contains a bond $b(s,s')$ joining them, a subset of active sites $C\subset\GG$ is called a {\em (connected) cluster}\index{(connected) cluster} if any two sites in $C$ can be connected by a chain of neighboring active sites. Thus, the notion of connectedness is determined by the configuration of bonds between sites in the graph, i.e. by the topology of $\GG$.\\

\noindent We will now concentrate on the random variable given by the {\em maximum cluster size}\index{maximum cluster size}
\begin{equation}\label{maxcluster}
M := \max \lbrace \vert C \vert\,:\, C\subset\GG \,\,\mathrm{cluster}\rbrace .
\end{equation}
The basic idea of the algorithm is now to split up the probabilities using the random variable
\begin{equation}\label{maxcluster2}
N := \vert \lbrace s\in \GG\,:\, s \,\,\mathrm{active}\rbrace \vert.
\end{equation}
given by the {\em number of active sites}\index{number of active sites}. Then
\begin{equation}\label{maxclusterdist}
\P(M = m) = \sum_{n=0}^S \P(M=m\,\vert\, N=n)\,\P(N=n)
\end{equation}
and since the sites are active {\em mutually independent} with probability $p$, the number of active sites is {\em binomially distributed}, i.e.
$$
\P (N=n) = \left(\begin{array}{c}S\\n\end{array}\right)\,p^n (1-p)^{S-n} .
$$
Furthermore, the conditional probabilities $p_{mn}:= \P(M=m\,\vert\, N=n)$ depend solely on the topology of the graph. The {\em activation probability} $p$ influences the distribution of $M$ via adjusting the number $N$ of active bonds alone.\\

\noindent The {\em Newman - Ziff algorithm}\index{Newman - Ziff algorithm} is an algorithm to simulate those conditional probabilities $p_{mn}$. Let for fixed $n$ denote $M_n\in\lbrace 0,1,...,n\rbrace$ be the random variable that is distributed by $p_{nm}$. Then we can consider all $M_n$ simultaneously as being generated by {\em random permutations}\index{random permutations} $\Pi$ of $\lbrace 1,...,S\rbrace$:\\

\noindent We consider a permutation $\pi$ as a {\em random visiting scheme}\index{random visiting scheme} of the sites of $\GG$ and $\pi(1),...,\pi(n)$ to be the first $n$ sites that are visited. Considering these sites as active, the set $\lbrace\pi(1),...,\pi(n)\rbrace$ consists of several connected clusters. We denote by $X_n(\pi)$ the size of the largest of these clusters.

\begin{Lemma} The distribution of $X_n(\pi)$ coincides with the one of $M_n$ for all $n$.
\end{Lemma}

\begin{Proof} A random permutation is given by $\P (\Pi = \pi) = 1/S !$, hence
$$
\P (\Pi(1)=\pi(1),...,\Pi(n)=\pi(n)) = \left(\begin{array}{c}S\\n\end{array}\right)^{-1}
$$
and all possible configurations of $n$ active sites are equi-distributed. This is the same if you choose active sites independently with probability $p$ and consider the conditional probability having $n$ active sites.
\end{Proof}

\noindent The advantage of this is twofold: First of all, we have to generate only one random permutation to get estimators $\widehat{M}_n$ for all $n=1,...,S$, and, crucially, we can add active sites inductively according to the random visiting scheme generated by the permutation which makes it much easier and less storage consuming to keep track of the largest cluster.\\

\noindent Thus, one single run of the estimation procedure consists of
\begin{enumerate}
\item let $\widehat{M}_0 = 0$,
\item generate a random permutation,
\item add active sites inductively according to the random visiting scheme generated by the permutation,
\item after adding site $n$, check whether this site leads to enlargement of one or the merging of several already existing clusters,
\item check whether the size $m$ of the new cluster generated by adding site $n$ is larger than $\widehat{M}_{n-1}$,
\item if so, let $\widehat{M}_n=m$, otherwise $\widehat{M}_n=\widehat{M}_{n-1}$,
\item record $\widehat{M}_n$, $n=1,...,S$.
\end{enumerate}
For practical implementations, the active sites are recorded by assigning a {\em label} according to which cluster the site belongs (0: not active, 1: belonging to cluster 1,...). These labels have to be updated every time a site is added. This requires some work if adding of a new site results in merging several different clusters. Thus, in the course of one single run (meaning a run according to one generated random permutation), we also have to record the current {\em label vector}\index{label vector} $L \in \lbrace 0,1,...,S\rbrace^S$. The conditional probabilities can then be estimated by carrying out several runs of this scheme and using the strong law of large numbers for suitably chosen indicator functions as usual. This holds for the conditional probabilities as well as for the unconditional distribution of the maximum cluster size after convoluting with the binomial distribution of the number of active sites above.\\

\subsection{The modified Newman - Ziff algorithm}\label{MNZ}


To simulate the type II error for the case of small images, in \cite{langovoy_report_Robust_Detection} we have used a modified version of the Newman - Ziff algorithm suitable for site occupation probabilities which are inhomogeneous over the lattice. The modified Newman - Ziff algorithm computes the distribution of the magnitudes of the largest clusters when the site occupation probability is inhomogeneous.

To be precise, we have a subgrid $\GG'\subset \GG$ where the occupation probability is $p_1$, whereas it is $p_2$ for sites in $\GG - \GG'$. In that case, we have to store the largest cluster sizes for all configurations $(N,M) = (n,m)$ of $n\leq \vert\GG'\vert$ occupied sites in $\GG'$ and $m\leq\vert\GG\vert$ occupied sites in $\GG$. The probabilities of those configurations are
$$
\P((N,M) = (n,m)) = \left(\begin{array}{c}\vert\GG'\vert\\n\end{array}\right)\left(\begin{array}{c}\vert\GG\vert\\m\end{array}\right)p_1^n(1-p_1)^{\vert\GG'\vert - n}p_2^m(1-p_2)^{\vert\GG\vert - m}.
$$
The basic idea of the algorithm remains essentially the same. We solved the additional problem of running through all configurations by using two different visiting schemes, one for inside $\GG'$ and another one for outside $\GG'$.


\subsection{R-Code: Newman - Ziff algorithm}\label{NZSIM}

The implementation consists of one main function {\tt simulation} which produces a list of simulated distribution functions of the maximum cluster size for a rectangular $n\times m$-grid with six-neighborhood for different activation probabilities $p$ ($n$ and $m$, the number of runs {\tt runs} and the different activation probabilities {\tt probs} under consideration are specified in the header and have to be changed manually).\\

\noindent Furthermore, we have six subroutines, given by
\begin{enumerate}
\item {\tt distribution} which produces an empirical distribution function of the maximum cluster size out of the different estimators $(\widehat{M}_0,...,\widehat{M}_n)$ produced by the subsequent runs of {\tt run},
\item {\tt generate} generates a suitable vector of binomial probabilities,
\item {\tt run} one run of the Newman - Ziff algorithm as described above, generates the estimators $(\widehat{M}_0,...,\widehat{M}_n)$,
\item {\tt rvsch} generates the random visiting scheme,
\item {\tt sixngh} is our way to implement the neighborhood structure (i.e. the topology of the graph),
\item {\tt sizeprob} generates an unconditional empirical distribution function.
\end{enumerate}

\begin{verbatim}
############################ MAIN #########################
##
## 'simulation'
##
## uses: 'distribution'
##
## change 'n','m','runs','probs' manually

simulation <- function()
                {
                n <- 55
                m <- 55
                runs <- 1000

                probs <- c(0.1,0.2,0.3,0.4,0.42,0.44,0.46,
                           0.48,0.5,0.52,0.54,0.56,0.58,
                           0.6,0.7,0.8,0.9)

                N <- n*m
                all_dist <- list()

                for(PP in 1:17)
                {
                all_dist[[PP]] <- distribution(n,m,runs,probs[PP])
                print(PP)
                }

                return(all_dist)
                }


##########
##
##  'distribution' generates estimate of cdf
##
##  uses: 'generate','sixngh','sizeprob','run'
##  used by: 'simulation'
##

distribution <- function(n,m,runs,p)
{
N <- n*m

ngh <- sixngh(n,m)
probs <- rep(0,N)

binprob <- generate(N,p)

for(r in 1:runs)
{
size <- run(n,m,ngh)
sizeprob <- sizeprob(size,binprob)

probs <- probs + sizeprob
}

probab <- probs/runs
return(probab)
}


##########
##
## 'generate' generates vector of binomial probabilities
##
## input: n 'number of trials', p 'success probability'
## output: Bin(n,p) probability vector
## used by: distribution

generate <- function(n,p){outcomes <- c(0:n)
                          probs <- dbinom(outcomes,n,p)
                          return(probs)}


##########
##
## 'run' -- "one single run of the newman ziff algorithm"
##
## remark: the 'updating step' uses special properties of
## six-neighborhoods for the conditioning

## input: n,m 'grid rows and columns'
##        ngh 'neighborhood structure, provided by 'sixngh'
## uses: 'sixngh', 'rvsch'
## output: size 'size[TT]: maximum cluster size for TT
## occupied bonds'

run <- function(n,m,ngh)
{
## define vector 'size' and set N 'number of edges'
size <- c()
N <- n*m

## generate VS <- rvsch(N) 'random visiting scheme'
## == random permutation of 1,...,N
VS <- rvsch(N)

## initialise clus 'cluster label vector'
## 0: site not occupied, k>0: site belongs to cluster k
clus <- rep(0,N)

## T 'step number' (for a single run)

for(TT in 1:N)
    {
    ## maximal cluster label
      maxi <- max(clus)

    ## random edge
      edge <- VS[TT]

    ## neighbors of random edge
      neig <- ngh[[edge]]

    ## all cluster labels in neighborhood of random edge,
    ## ordered by magnitude
    ## lbl = 0: unoccupied edge, lbl = k: edge belongs to
    ## cluster k
    ## remark: a moment of reflection shows that always
    ## length(labl)<=4 for the triangular lattice
      labl <- sort(unique(clus[neig]))
      L <- length(labl)

    ## updating labels:
    ## case #1: no occupied bonds in the neighborhood
    ## case #2: random edge connects all clusters in the
    ##          neighborhood, new label = maximum label

    if(labl[L] == 0){clus[edge] <- 1 + maxi}else
        {
        if(L == 1){clus[edge] <- labl[1]}else
        {
        for(k in 2:L)
        {
        z <- (labl[L] - labl[k])*(1-abs(sign(clus - labl[k])))
        clus <- clus + z
        clus[edge] <- labl[L]
        }
        }
        }

    ## calculate the maximum cluster size
    ## either the cluster just constructed by merging the clusters
    ## in the neighborhood of the random edge is now the largest
    ## cluster, or the largest cluster remains the same

    newlabel <- clus[edge]
    mergedsize <- length(clus[clus == newlabel])

     if(TT > 1)
     {
     if(mergedsize > size[TT-1]){size[TT]<-mergedsize
     }else{size[TT]<-size[TT-1]}
     }
     else{size[1] <- 1}
    }

## output 'size' vector
return(size)
}


##########
##
## 'rvsch'  "random visiting scheme"
##
## used by: 'run'

rvsch <- function(N){ x <- c(1:N)
                      z <- sample(x,N)
                      return(z)
                    }

##########
##
## 'sixngh' 6-neighborhood (triangular lattice)
##
## used by: 'run'

sixngh <- function(n,m)
{

neighbors <- list()

## corners

neighbors[[1]]           <- c(2,n+1)
neighbors[[n]]           <- c(n-1,2*n,2*n-1)
neighbors[[n*m]]         <- c(n*(m-1),n*m -1)
neighbors[[(m-1)*n + 1]] <- c((m-1)*n + 2,(m-2)*n+1,(m-2)*n+2)

## boundary

# (a) above
for(i in 2:(n-1)){neighbors[[i]]<- c(i-1,i+1,i+n,i+n-1)}

# (b) below
for(i in ((m-1)*n+2):(n*m-1)){neighbors[[i]]<- c(i-1,i+1,i-n,i-n+1)}

### (c) left
for(i in 1:(m-2))
{neighbors[[n*i+1]] <- c(n*(i-1)+1,n*(i-1)+2,n*(i+1)+1,n*i+2)}

### (d) right
for(i in 2:(m-1)){neighbors[[n*i]]<- c(n*(i-1),n*(i+1),n*(i+1)-1,n*i-1)}

## interior
for(i in 1:(m-2))
{
for(j in 1:(n-2))
{
neighbors[[i*n + 1 + j]]<- c(i*n + j,i*n + 2 + j,(i-1)*n+1+j,
                            (i-1)*n+j+2,(i+1)*n+1+j,  (i+1)*n+j)
}
}
return(neighbors)
}


##########
##
## 'sizeprob' generates unconditional empirical cdf
##
## used by: 'distribution'
## input: k cluster size, integer
##        'binprob' suitable binomial probabilities from 'generate'
##        'size' vector from simulation 'run'
##
## output: 'estimator' vector of P(max <= k), k = 0,...,n

sizeprob <- function(size,binprob)
        {
        estimator <- c()
        size_ext <- c(0,size)

        for(k in 1:length(size))
        {
        ## 'indicator' determines all entries of 'size_ext'
        ## less or equal than k
        indicator <- 0.5*(1 - sign(size - k - 0.5))

        ## 'estimator' is the estimator of P(max <= k) based
        ## on the single run
        estimator[k] <- sum(c(1,indicator) * binprob)
        }
        return(estimator)
        }
\end{verbatim}

\subsection{R-Code: modified Newman - Ziff algorithm}\label{MNZSIM}


The modified algorithm uses the functions {\tt sixngh} and {\tt rvsch} from the previous collection and furthermore the following code.

\begin{verbatim}
########################## MAIN_MODI ##############################################
##
## 'simulation_modi'
##
## uses: 'run_modi','sizeprob_modi'
##
## change manually: x,y gridsize, runs number of runs, s_set subgrid,
##                  P_in, P_out site occupation probabilities inside and
##                  outside the subgrid
##
## output:          'estimate' estimated cluster size distribution


########################## MAIN_MODI ##################################
##
## 'simulation_modi'
##
## uses: 'run_modi','sizeprob_modi'
##
## change manually: x,y gridsize, runs number of runs, s_set subgrid,
##                  P_in, P_out site occupation probabilities inside and
##                  outside the subgrid
##
## output:          'estimate' estimated cluster size distribution


simulation_modi <- function()
{
## we consider fixed grid size (square grid) with 55x55 sites

x <- 55
y <- 55
n <- x*y
runs <- 100

## 's_set' is a vector giving the positions of sites belonging to the subset
## positions are given by a vector, the sites in the grid are labelled as:
##
##   1,2,3,...,x
##   x+1,...., 2x
##      ......
##  (y-1)x+1... xy
##

## example: 10x10-subset left upper corner at location (20,20)

s_set <- c(c(1065:1074),c(1120:1129),c(1175:1184),c(1230:1239),c(1285:1294),
c(1340:1349),c(1395:1404),c(1450:1459),c(1505:1514),c(1560:1569))
ngh <- sixngh(x,y)
P_in <- 0.6
P_out <- 0.4

est <- rep(0,n)

for(LL in 1:runs)
{
single_sim <- run_modi(x,y,s_set,ngh)
est <- est + sizeprob_modi(single_sim,P_in,P_out)
print(LL)
}

estimate <- est/runs

return(estimate)
}


## function "run_modi" -- 'one single run of the modified newman ziff algorithm'
## remark: the 'updating step' uses special properties of six-neighborhoods for
## the conditioning

## input: n,m 'grid rows and columns'
##        ngh 'neighborhood structure, provided by 'sixngh'
##        s_v 'positions of subset sites'
## functions used: 'sixngh', 'rvsch'
##
## output:  size_out (list) 'size[[a]][b]: maximum cluster size for a-1 occupied bonds inside the subset
##          and b-1 occupied bonds outside the subset'

run_modi <- function(n,m,s_v,ngh)
{
## define vector 'size' and set M 'number of edges in the subset' and
## N number of sites in the grid
size <- c()
M <- length(s_v)
N <- n*m

## generate VS <- rvsch(M) 'random visiting scheme of subset'
## == random permutation of 1,...,M
VS <- rvsch(M)

## initialise clus 'cluster label vector'
## 0: site not occupied, k>0: site belongs to cluster k
clus <- rep(0,N)

## initialize list of all cluster vectors

s_clus <- list()

## TT 'step number' (for a single run for the subset)

for(TT in 1:M)
    {
    ## maximal cluster label
      maxi <- max(clus)

    ## random edge inside subset
      edge <- s_v[VS[TT]]

    ## neighbors of random edge
      neig <- ngh[[edge]]

    ## all cluster labels in neighborhood of random edge, ordered by magnitude
    ## lbl = 0: unoccupied edge, lbl = k: edge belongs to cluster k
    ## remark: a moment of reflection shows that always length(labl)<=4 for the
    ## triangular lattice
      labl <- sort(unique(clus[neig]))
      L <- length(labl)

    ## updating labels: case #1: no occupied bonds in the neighborhood
    ##                  case #2: random edge connects all clusters in the
    ##                           neighborhood, new label = maximum label

    if(labl[L] == 0){clus[edge] <- 1 + maxi}else
        {
            if(L == 1){clus[edge] <- labl[1]}else
                {
                  for(k in 2:L)
                      {
                        z <- (labl[L] - labl[k])*(1-abs(sign(clus - labl[k])))
                        clus <- clus + z
                        clus[edge] <- labl[L]
                      }
                }
        }

    ## calculate the maximum cluster size
    ## either the cluster just constructed by merging the clusters in the
    ## neighborhood of the random edge is now the largest cluster, or the
    ## largest cluster remains the same

    newlabel <- clus[edge]
    mergedsize <- length(clus[clus == newlabel])

     if(TT > 1)
     {
     if(mergedsize > size[TT-1]){size[TT]<-mergedsize}else{size[TT]<-size[TT-1]}
     }
     else{size[1] <- 1}

    ## cluster status vector 'clus' has now to be stored for all numbers of
    ## occupies sites in the subset

    s_clus[[TT]] <- clus
    }


## output 'status' vector 's_clus' + value for 0 active subset sites, stored in
## s_clus[[M+1]], 'size' == 0

s_clus[[M+1]] <- rep(0,N)
size[M + 1] <- 0

## second part: adding active sites to the complement of the subset
## remark: the 'updating step' uses special properties of six-neighborhoods for
## the conditioning

## define vector 'o_v': complement of subset
latt <- c(1:N)
o_v <- setdiff(latt,s_v)
J <- length(o_v)
anoth_size <- c()

## generate VS <- rvsch(L) 'random visiting scheme'
## of the complement
WS <- rvsch(J)

## initialize list of all cluster sizes
s_clussize <- list()

## WW 'step number' for initial subset configuration
VV <- M+1

for(WW in 1:VV)
{
clus <- s_clus[[WW]]

## TT 'step number' (for a single run)

for(TT in 1:J)
    {
    ## maximal cluster label
      maxi <- max(clus)

    ## random edge inside subset
      edge <- o_v[WS[TT]]

    ## neighbors of random edge
      neig <- ngh[[edge]]

    ## all cluster labels in neighborhood of random edge, ordered by magnitude
    ## lbl = 0: unoccupied edge, lbl = k: edge belongs to cluster k
    ## remark: a moment of reflection shows that always length(labl)<=4 for the
    ## triangular lattice
      labl <- sort(unique(clus[neig]))
      L <- length(labl)

    ## updating labels: case #1: no occupied bonds in the neighborhood
    ##                  case #2: random edge connects all clusters in the
    ##                           neighborhood, new label = maximum label

    if(labl[L] == 0){clus[edge] <- 1 + maxi}else
        {
            if(L == 1){clus[edge] <- labl[1]}else
                {
                  for(k in 2:L)
                      {
                        z <- (labl[L] - labl[k])*(1-abs(sign(clus - labl[k])))
                        clus <- clus + z
                        clus[edge] <- labl[L]
                      }
                }
        }

    ## calculate the maximum cluster size
    ## either the cluster just constructed by merging the clusters in the
    ## neighborhood of the random edge is now the largest cluster, or the
    ## largest cluster remains the same

    newlabel <- clus[edge]
    mergedsize <- length(clus[clus == newlabel])

     if(TT > 1)
     {
     if(mergedsize > anoth_size[TT-1])
     {anoth_size[TT] <- mergedsize}else{anoth_size[TT] <- anoth_size[TT-1]}
     }
     else{
          if(mergedsize > size[WW])
          {anoth_size[1] <- mergedsize}else{anoth_size[1] <- size[WW]}
         }
    }

s_clussize[[WW]] <- anoth_size
}

## real output 'size_out'

size_out <- list()
size_out[[1]] <- c(0,s_clussize[[M+1]])
for(l in 1:M)
{
size_out[[l+1]] <- c(size[l],s_clussize[[l]])
}

return(size_out)
}


## 'sizeprob_modi'  calculates an estimator for the probability of maximum
##                  cluster sizes not to exceed a given value k
##
## input:   'P_in', 'P_out' site occupation probabilities inside and outside
##          the subset
##          'size_out' list from simulation 'run_modi'
##
## output: 'estimator' vector of P(max <= k), k = 0,...,n

sizeprob_modi <- function(size_out,P_in,P_out)
        {
        indicator <- list()
        estimator <- c()

        # number of inner sites
        VV <- length(size_out)
        V <- VV - 1
        # number of outer sites
        WW <- length(size_out[[1]])
        W <- WW - 1
        # total number of sites
        NN <- V + W

        # binomial probability vectors
        out_sites <- c(0:W)
        bin_out <- dbinom(out_sites,W,P_out)

        in_sites <- c(0:V)
        bin_in <- dbinom(in_sites,V,P_out)


        for(k in 1:NN)
        {
        estimator[k] <- 0
        for(i in 1:VV)
        {
        ## 'indicator' determines all entries of 'size_out' les
        s or equal than k
        indicator[[i]] <- 0.5*(1 - sign(size_out[[i]] - k - 0.5))

        ## 'estimator' is the estimator of P(max <= k) based on the single run
        estimator[k] <- estimator[k] + bin_in[i] * sum(indicator[[i]]*bin_out)
        }
        }
        return(estimator)
}
\end{verbatim}

\smallskip
\noindent {\bf Acknowledgments.} The authors would like to thank Laurie Davies and Alessandro Di Bucchianico for helpful discussions. \\

\bibliographystyle{plainnat}

\bibliography{papiere}


\end{document}